\documentclass[english]{beamer}
\usepackage{mathptmx}
\usepackage[T1]{fontenc}
\usepackage[latin9]{inputenc}
\usepackage{amsmath}
\usepackage{amssymb}

\makeatletter
 \newcommand\makebeamertitle{\frame{\maketitle}}%
 \AtBeginDocument{
   \let\origtableofcontents=\tableofcontents
   \def\tableofcontents{\@ifnextchar[{\origtableofcontents}{\gobbletableofcontents}}
   \def\gobbletableofcontents#1{\origtableofcontents}
 }
 \long\def\lyxframe#1{\@lyxframe#1\@lyxframestop}%
 \def\@lyxframe{\@ifnextchar<{\@@lyxframe}{\@@lyxframe<*>}}%
 \def\@@lyxframe<#1>{\@ifnextchar[{\@@@lyxframe<#1>}{\@@@lyxframe<#1>[]}}
 \def\@@@lyxframe<#1>[{\@ifnextchar<{\@@@@@lyxframe<#1>[}{\@@@@lyxframe<#1>[<*>][}}
 \def\@@@@@lyxframe<#1>[#2]{\@ifnextchar[{\@@@@lyxframe<#1>[#2]}{\@@@@lyxframe<#1>[#2][]}}
 \long\def\@@@@lyxframe<#1>[#2][#3]#4\@lyxframestop#5\lyxframeend{%
   \frame<#1>[#2][#3]{\frametitle{#4}#5}}
 \newenvironment{topcolumns}{\begin{columns}[t]}{\end{columns}}
 \def\lyxframeend{} 

\setbeamercovered{transparent}

\makeatother

\usepackage{babel}
\begin{document}

\title[Predicting compressible sequences]{Two betting strategies that predict all compressible sequences}

\author{T.~Petrovi\'{c}\inst{} }

\institute[tomeepx@gmail.com]{\inst{}tomeepx@gmail.com}

\date[CCR 2012]{Seventh International Conference on Computability, Complexity and
Randomness (CCR 2012)}

\makebeamertitle




\lyxframeend{}\lyxframe{Outline}

\tableofcontents{}

\lyxframeend{}\section{Betting Games}

\lyxframeend{}\subsection[Sequence-set Betting]{Definition of Sequence-set Betting }

\lyxframeend{}\lyxframe{Definition of Sequence-set Betting}

\framesubtitle{(using mass-distribution)}

\begin{columns}

\column{7cm}
\begin{itemize}
\item <1->Start with$(S,m)$
\item <1->Betting decision $(S_{0},m_{0})$ , $(S{}_{1},m_{1})$ \\$m_{0}+m_{1}=m$\\$\lambda(S_{0})=\lambda(S_{1})=\frac{1}{2}\lambda(S)$ 
\item <1->Sequence in $S_{0}$, next is$(S_{0},m_{0})$\\otherwise start
with $(S_{1},m_{1})$ 
\item <1->Success: capital $c=m/\lambda(S)$\\rises unboundedly
\item <2->Decision tree descrbes betting strategy
\end{itemize}

\column{3cm}
\begin{overprint}
\includegraphics<1>[scale=0.15]{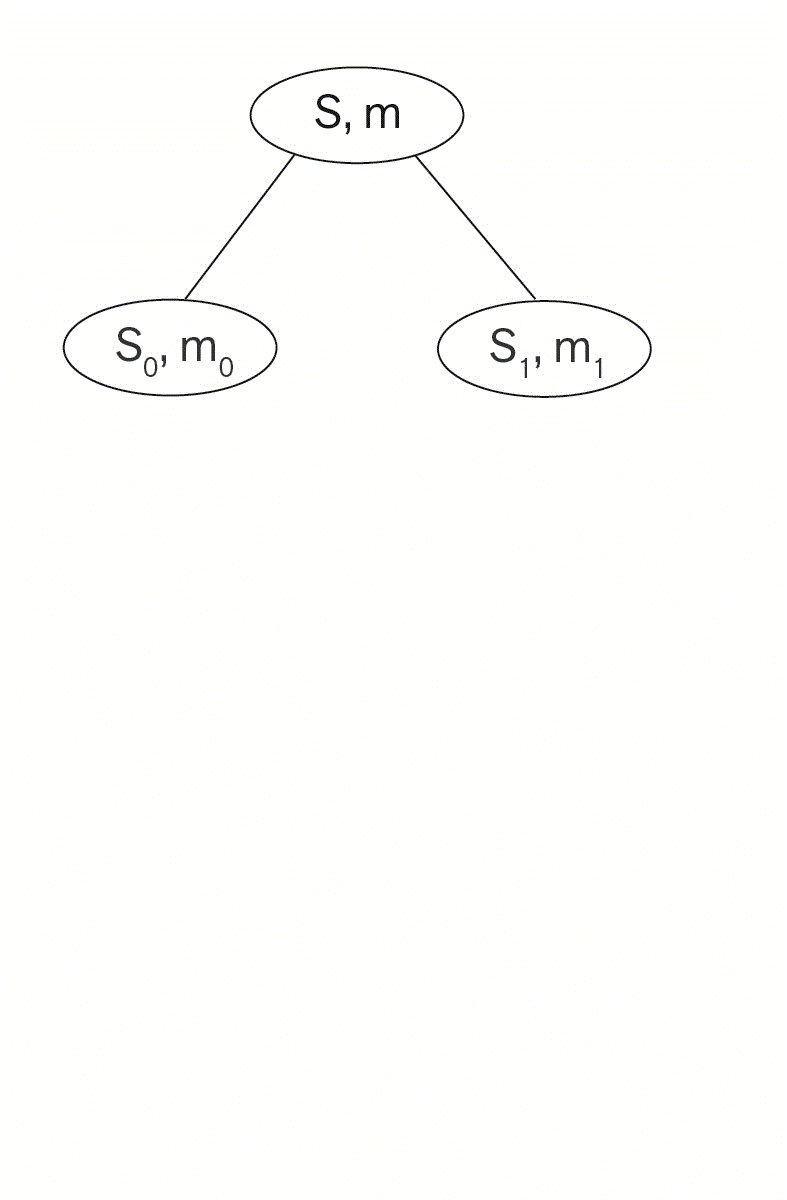}
\includegraphics<2->[scale=0.15]{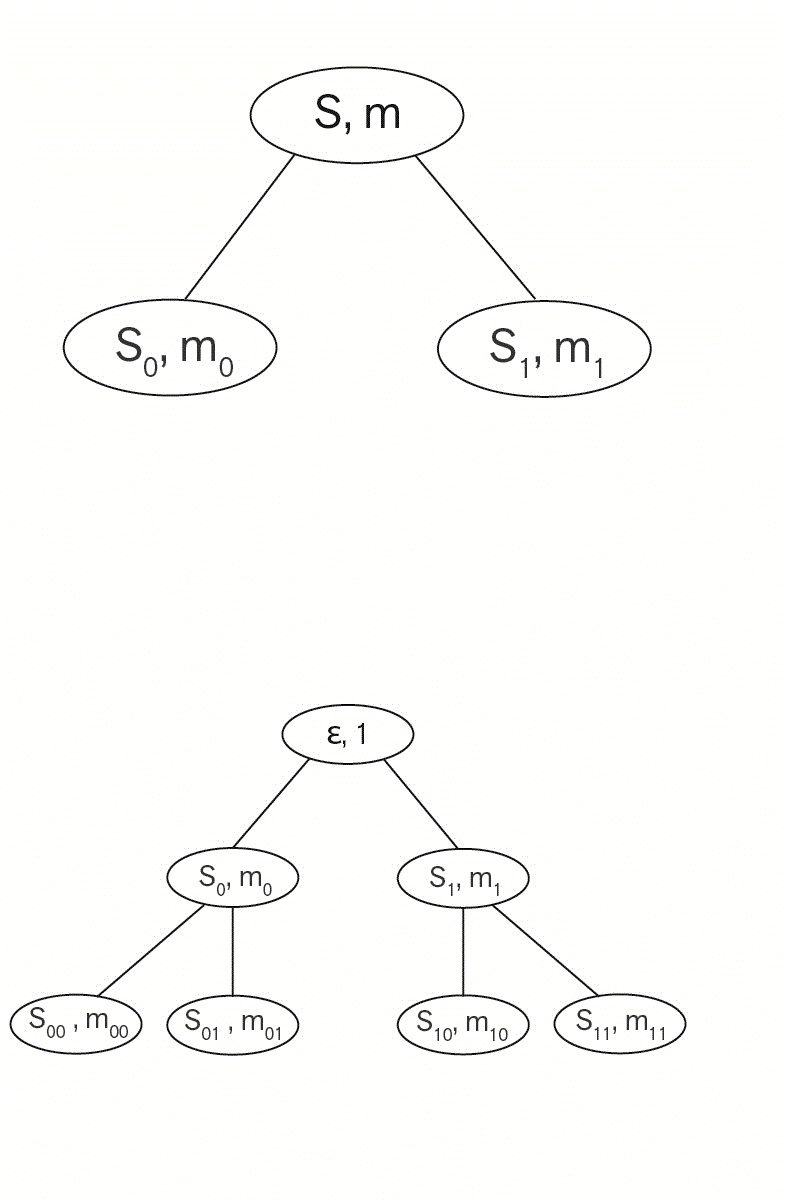}
\end{overprint}
\end{columns}

\lyxframeend{}\subsection{Comparison with Martingale Processes}

\lyxframeend{}\lyxframe{Martingale Processes }

\begin{columns}

\column{7cm}
\begin{itemize}
\item <1->Function $d$ from words to reals
\item <1->Equivalence relation on words $\approx_{d}$\\$v\approx_{d}w$
if for \\$v^{\prime}\preceq v,\: w^{\prime}\preceq w,\: l(v^{\prime})=l(w^{\prime})$\\we
have $d(v^{\prime})=d(w^{\prime})$
\item <1->Fairness condition:\\$\qquad$$2\underset{\{v:v\approx_{d}w\}}{\sum}d(v)$\\$=\underset{\{v:v\approx_{d}w\}}{\sum}[d(v0)+d(v1)]$
\end{itemize}

\column{3cm}
\begin{overprint}
\includegraphics<1->[scale=0.15]{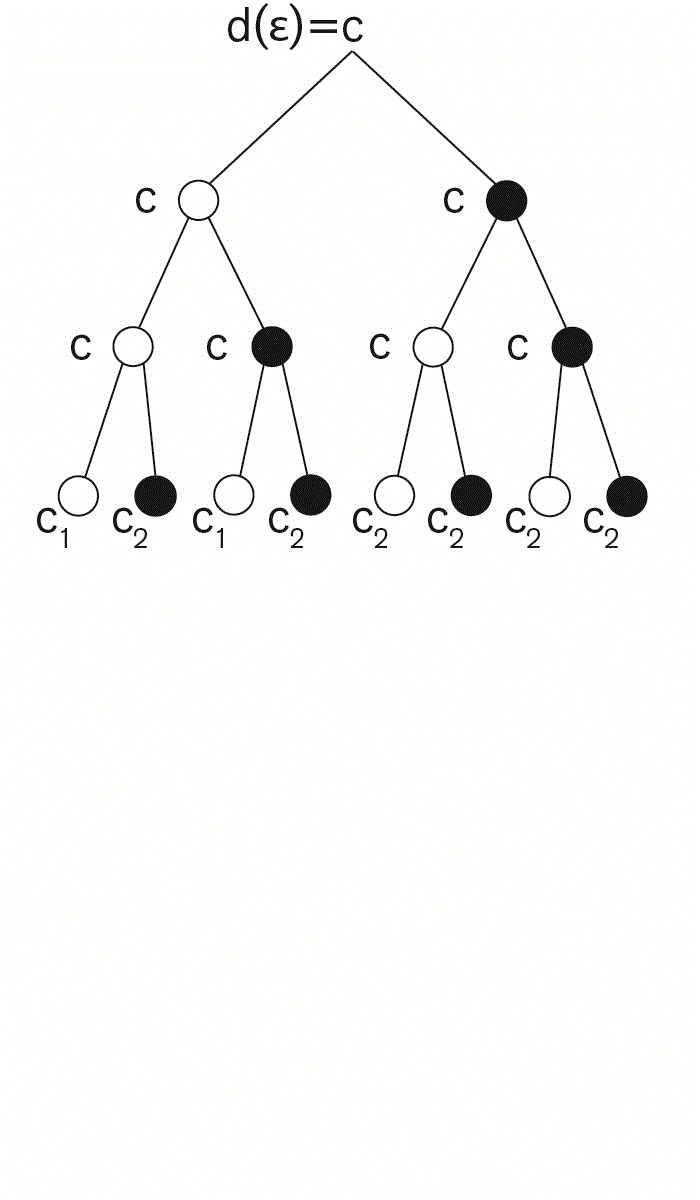}
\end{overprint}

\end{columns}

\lyxframeend{}\lyxframe{Modified Sequence-set Betting}

\begin{columns}

\column{7cm}
\begin{itemize}
\item <1->Replace $\lambda(S_{0})=\lambda(S_{1})=\frac{1}{2}\lambda(S)$\\with
$\lambda(S_{0})+\lambda(S_{1})=\lambda(S)$ 
\item <1->Betting game equivalent to martingale processes
\end{itemize}

\column{3cm}
\begin{overprint}
\includegraphics<1->[scale=0.15]{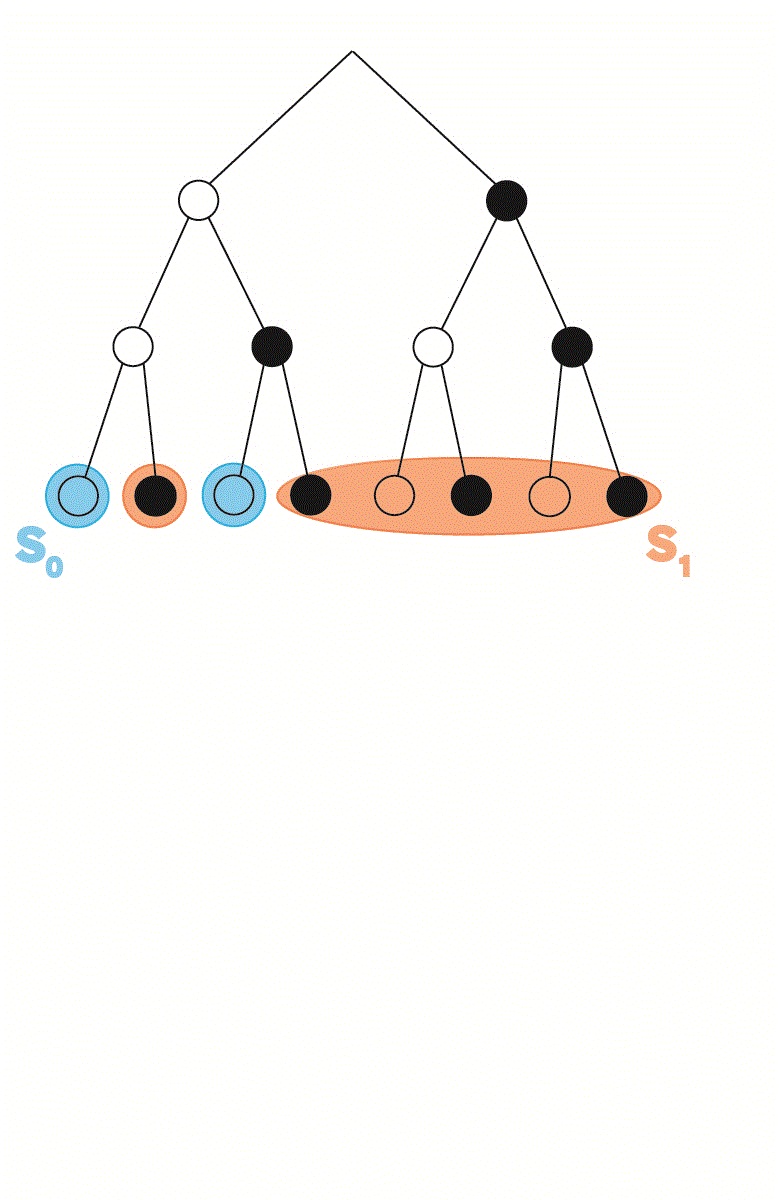}
\end{overprint}
\end{columns}
\begin{topcolumns}

\end{topcolumns}

\lyxframeend{}\lyxframe{Betting Strategy from Martingale Process}

\begin{columns}

\column{7cm}
\begin{itemize}
\item <1->$(S,m)\sim(V,d)$ , $V=\{v:v\approx_{d}w\}$\\$S\prec\alpha\Longleftrightarrow V\prec\alpha$
and $m=d(w)\lambda(V)$
\item <1->Find $v^{\prime}$ s.t. $v\prec v^{\prime},\; d(v^{\prime})\neq d(v)$
\item <1->$d$ divides $V$ into $V_{1},\ldots,V_{n}$
\item <1->Make according betting decisions
\item <1->No such $v^{\prime}$, make no further bets
\end{itemize}

\column{3cm}
\begin{overprint}
\includegraphics<1->[scale=0.15]{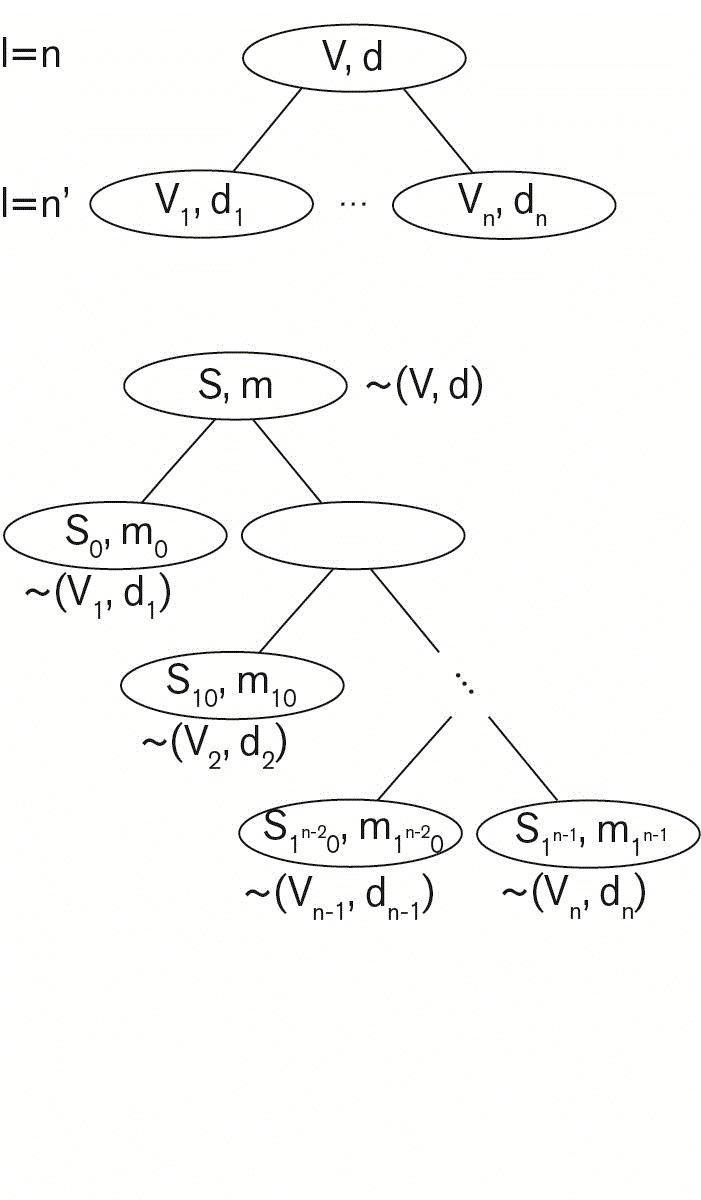}
\end{overprint}
\end{columns}
\begin{topcolumns}

\end{topcolumns}

\lyxframeend{}\lyxframe{Martingale Process from Betting Strategy}

\begin{columns}

\column{7cm}
\begin{itemize}
\item <1->$(S,m)\sim(V,d)$ , $V=\{v:v\approx_{d}w\}$\\$S\prec\alpha\Longleftrightarrow V\prec\alpha$
and $m=d(w)\lambda(V)$
\item <1->Betting decision $(S_{0},m_{0})$ , $(S_{1},m_{1})$ 
\item <1->Find \textrm{$n'$ $n'>l(w),\; n'\geq max(l(w):w\in S_{0}\cup S_{1}\}$}
\item <1->Set $d(v')=m_{0}/\lambda(S_{0})$ for $S_{0}\preceq v'$\\and
$d(v')=m_{1}/\lambda(S_{1})$ for $S_{1}\preceq v'$\\where $l(v')=n'$ 
\item <1->No betting decision for $(S,m)$~\\Set $d(v^{\prime})=d(w)$
for $V\prec v^{\prime}$
\end{itemize}

\column{3cm}
\begin{overprint}
\includegraphics<1->[scale=0.15]{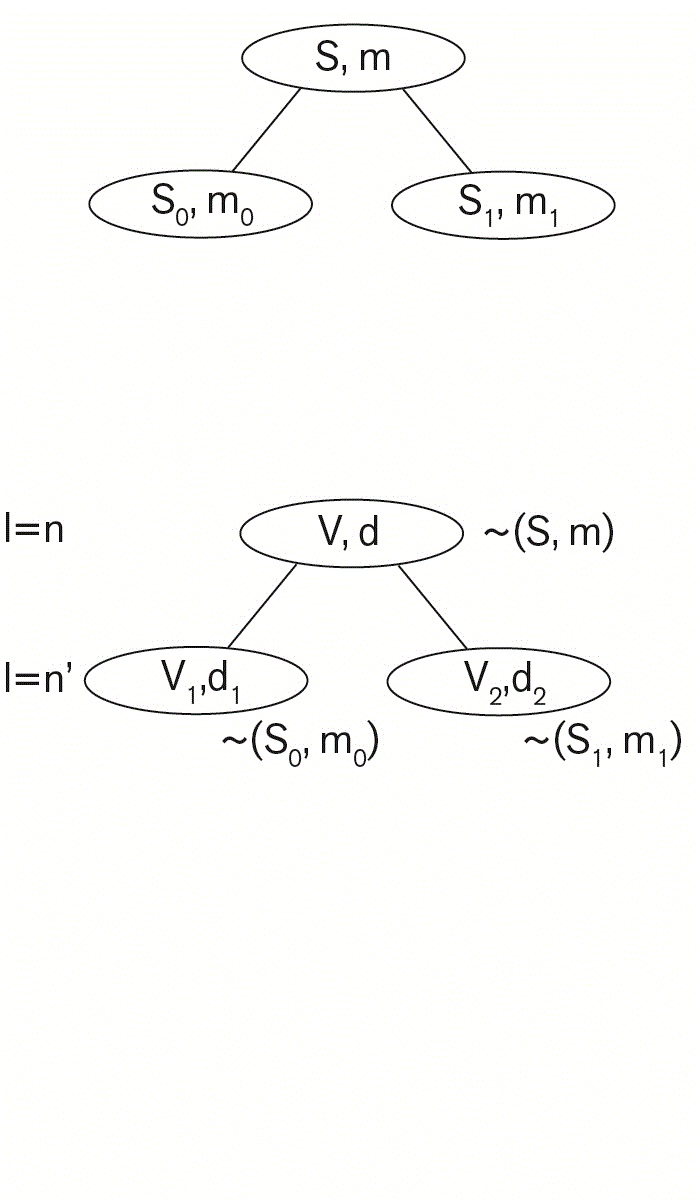}
\end{overprint}
\end{columns}
\begin{topcolumns}

\end{topcolumns}

\lyxframeend{}\subsection{Comparison with Non-monotonic Betting}

\lyxframeend{}\lyxframe{Comparison with Non-monotonic Betting}

\begin{columns}

\column{7cm}
\begin{itemize}
\item <1->Add to req. $\lambda(S_{0})=\lambda(S_{1})=\frac{1}{2}\lambda(S)$\\req.
that for some $n$\\$w\in S_{0}\Rightarrow w(n)=0$ $w\in S_{1}\Rightarrow w(n)=1$
\item <1->Betting decision $(S_{0},m_{0})$ , $(S_{1},m_{1})$\\places
a bet on bit value at position $n$
\end{itemize}

\note[item]{$m_{0}\geq m_{1}$ wager $w=(m_{0}-m_{1})/\lambda(P)$ on bit at
position $n$ being $0$}
\begin{itemize}
\item <1->Next iteration choses different position\\since previously picked
positions have\\all $0$'s or all $1$'s
\end{itemize}

\column{3cm}
\begin{overprint}
\includegraphics<1->[scale=0.15]{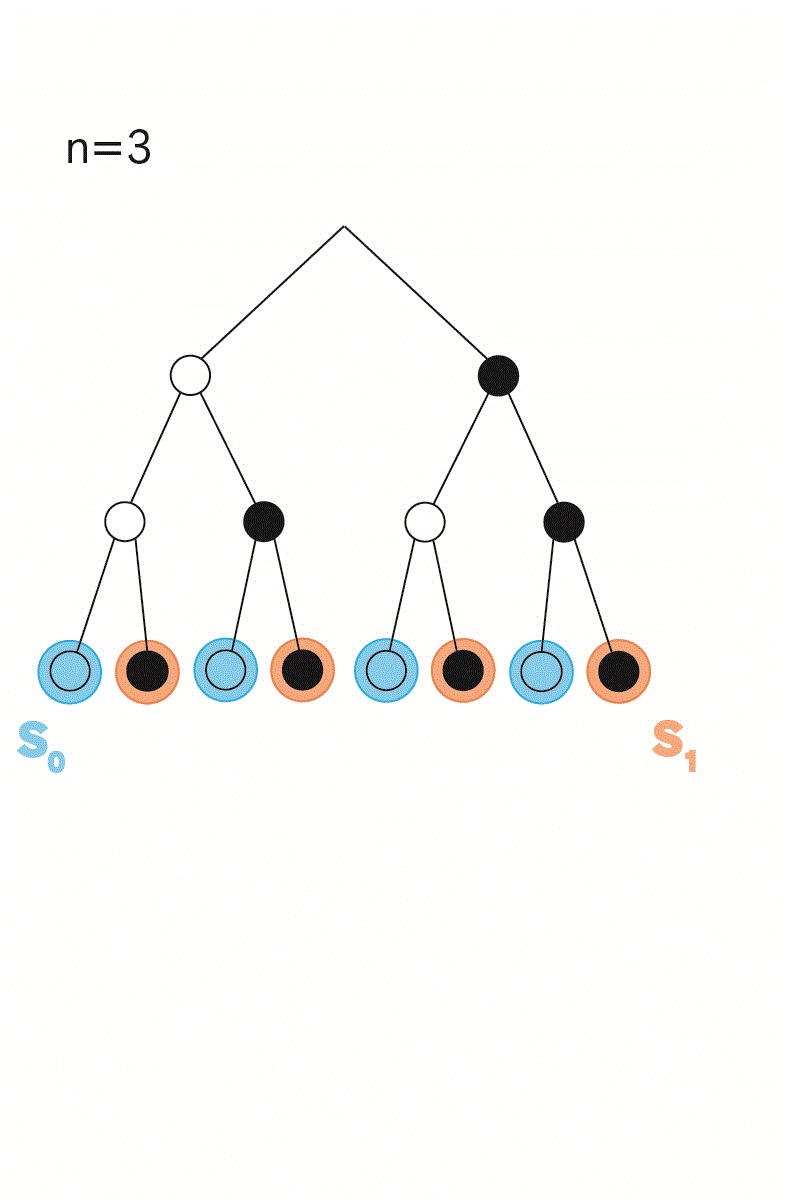}
\end{overprint}
\end{columns}
\begin{topcolumns}

\end{topcolumns}

\lyxframeend{}\lyxframe{Unpredictable Compressible Sequence}

\begin{columns}

\column{7cm}
\begin{itemize}
\item <1->Step1:For node $(S,m)$ wait for a bet
\item <1->No betting decision, strategy fails
\item <1->Choose node with less mass
\item <1->\textrm{If node small enough, compress all}
\item <1->\textrm{go to Step1}
\end{itemize}

\column{3cm}
\begin{overprint}
\includegraphics<1->[scale=0.15]{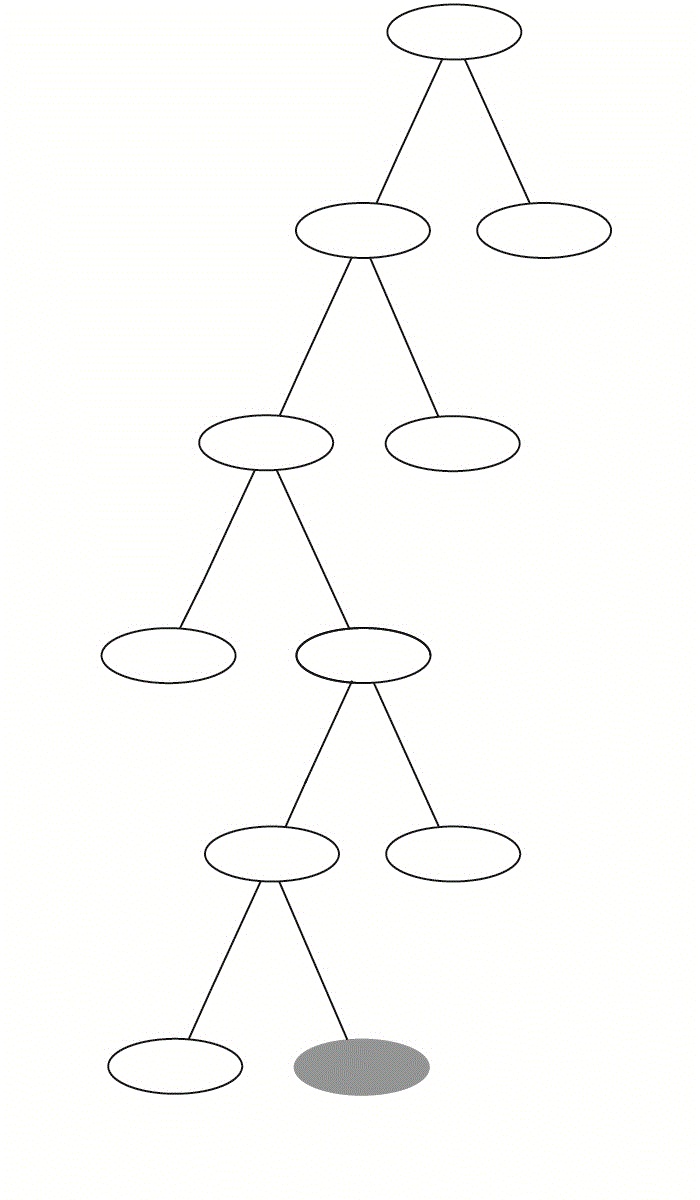}
\end{overprint}
\end{columns}

\global\long\def\sab{S_{b_{j}}^{a_{i}}}
\global\long\def\sai{S^{a_{i}}}
\global\long\def\sbj{S_{b_{j}}}

\lyxframeend{}\section{Algorithm that Constructs Two Betting Strategies}

\lyxframeend{}\lyxframe{Algorithm Outline}

\begin{itemize}
\item <1->\textrm{Algorithm constructs betting decision trees for strategies
$A$ and $B$}
\item <1->\textrm{The inputs are} some word $s$ and masses $ma,\: mb$
assigned to that prefix by \textrm{$A$ and $B$}
\item <1->Calculates $k$ from $(s,ma,mb)$
\item <1->Runs UTM to enumerate first prefixes $p$ that have inputs shorter
by $k$
\item <1->For each $p$ adds betting decisions, nodes that contain $p^{\prime},\: p\prec p^{\prime}$\\contain
only one word, same in \textrm{$A$ and $B$, for these start a new
instance of algorithm}
\item <1->If the set of prefixes is small, betting decisions such that
either\\\textrm{$A$ or $B$ double capital on that prefixes}
\end{itemize}

\lyxframeend{}\lyxframe{Notation}

\begin{overprint}
\includegraphics<1->[scale=0.15]{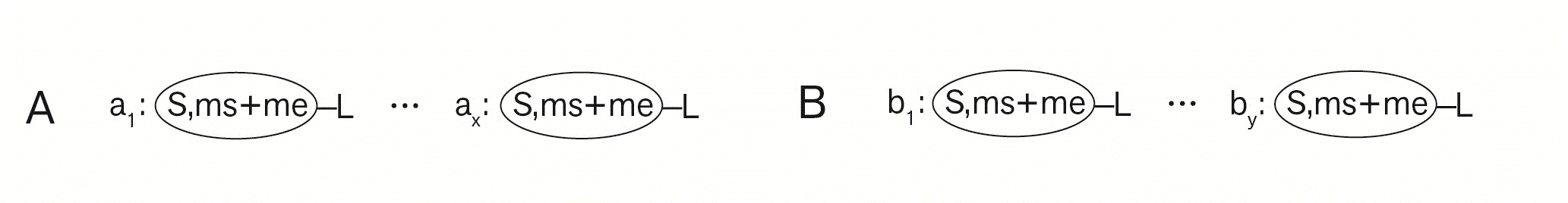}\end{overprint}
\begin{itemize}
\item <1->Only the leaf nodes of b.d.t. that don't contain sequences on
which another instance of algorithm was started are considered
\item <1->For each node $(a_{i},S,m)$, $(b_{j},S,m)$ two additional values
are used
\item <1->Mass reserved to ensure that no node has mass 0, $me$
\item <1->The portion of size of the so far found prefixes belonging to
the node $L$ 
\item <1->Mass $ms$ is used for doubling the capital on compressible prefixes,
mass assigned to node in b.d.t. $m=ms+me$
\item <1->\textrm{Denote indexes of considered leaf nodes of $A$ $a_{1},\ldots,a_{x}$,
of $B$ $b_{1},\ldots,b_{y}$}
\item <1->$(a_{i},S,ms,me,L)$ $\sai,ms^{a_{i}},me^{a_{i}},L^{a_{i}}$
\item <1->$(b_{j},S,ms,me,L)$ $\sbj,ms_{b_{j}},me_{b_{j}},L_{b_{j}}$
\end{itemize}

\lyxframeend{}\lyxframe{Initialization}

\begin{overprint}
\includegraphics<1->[scale=0.15]{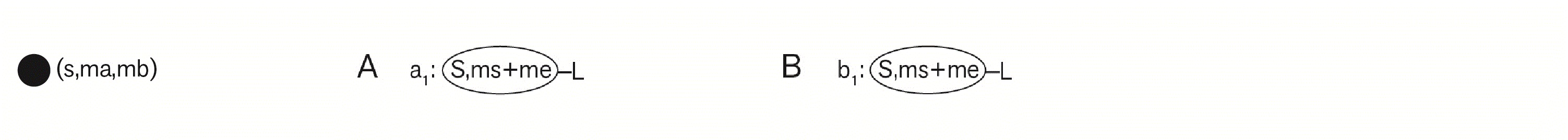}\end{overprint}
\begin{itemize}
\item <1->The input for the algorithm instance is $(s,ma,mb)$
\item <1->Set mass used for doubling the capital $m=min(ma,mb)/2$
\item <1->Set for $A$ \textrm{$S^{a_{1}}=\{s\},\: ms^{a_{1}}=m,\: me^{a_{1}}=ma-m,\: L^{a_{1}}=0$\\}$\;\quad$
for $B$ $S_{b_{1}}=\{s\},\: ms_{b_{1}}=m,\: me_{b_{1}}=mb-m,\: L_{b_{1}}=0$
\item <1->The capital for compressible prefixes will be $c=4(ma+mb)/\lambda(s)$
\item <1->set $k$ such that $2^{-k}<m^{2}(1-cs)/2\lambda(s)c^{2}(1+cs)^{2}$ 
\item <1->The $cs$ is some constant $0<cs<1$ , \textrm{in all iterations
$\sai\cap\sbj=\{\}$ \\ or $\lambda(s)\lambda(\sai\cap\sbj)$ is
between $(1\pm cs)(\lambda(\sai)+L^{a_{i}})(\lambda(\sbj)+L_{b_{j}})$}
\item <1->\textrm{The construction of b.d.t. starts by running the algorithm
for $(\epsilon,1,1)$}
\end{itemize}

\lyxframeend{}\lyxframe{Preparation Step}

\begin{columns}

\column{7cm}
\begin{itemize}
\item <1->Find next $p$, $\sai\cap\sbj\cap p\neq\{\}$
\item <1->If for all $a_{i},b_{j}$ $\sai\cap\sbj\cap p=\{\}$ or $\sai\cap\sbj\setminus p=\{\}$
skip rest
\item <2->For nodes $a_{i},b_{j}$ make $n$ betting decisions, $2^{n}$
new leaf nodes $a_{ig},b_{jh}$
\item <2->Evenly distribute mass and $L$
\item <2->Extend words not in intersection by $n$, distribute them ammong
leaf nodes
\item <2->Extend words in intersection and distribute them to have $S^{a_{ig}}\cap S_{b_{jh}}\cap p=\{\}$
or $S^{a_{ig}}\cap S_{b_{jh}}\setminus p=\{\}$
\end{itemize}

\column{3cm}
\begin{overprint}
\includegraphics<2->[scale=0.15]{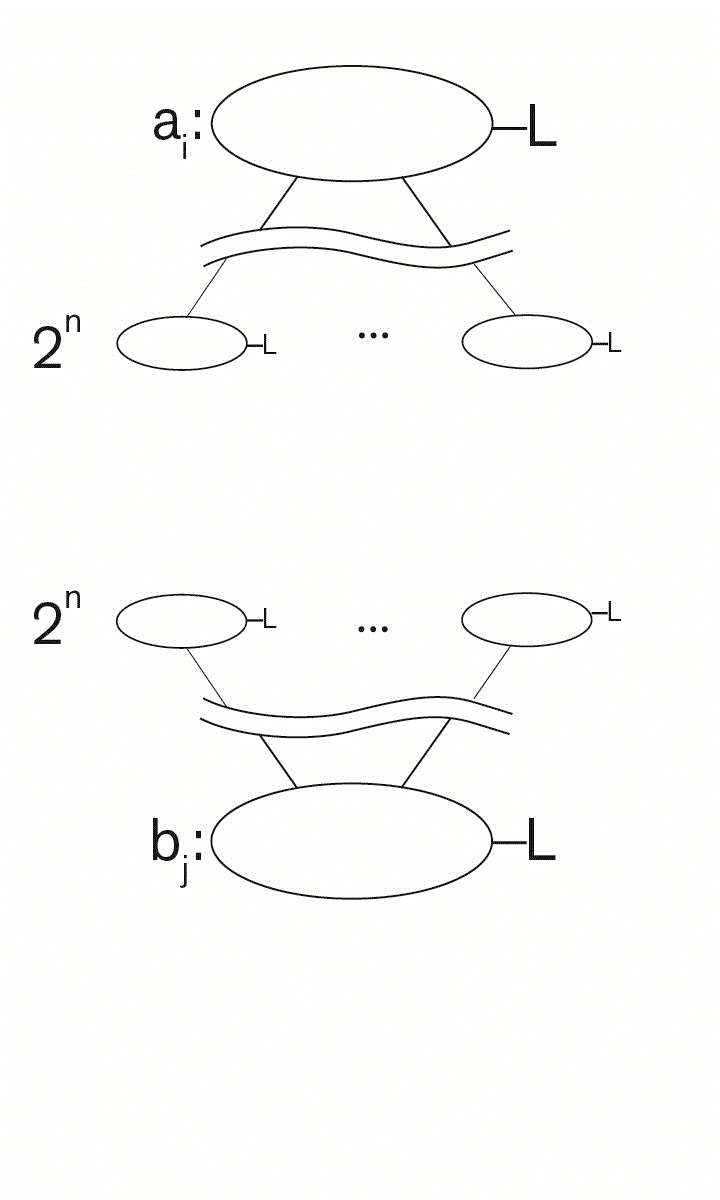}
\end{overprint}
\end{columns}
\vspace{0.1in}

\pause{If $\lambda(\sai\cap\sbj)=r(\lambda(\sai)+L^{a_{i}})(\lambda(\sbj)+L_{b_{j}})$
then $\lambda(S^{a_{ig}}\cap S_{b_{jh}})=r(\lambda(S^{a_{ig}})+L^{a_{ig}})(\lambda(S_{b_{jh}})+L_{b_{jh}})$}

\lyxframeend{}\lyxframe{Mass Assignment Step}

\begin{overprint}
\includegraphics<1->[scale=0.15]{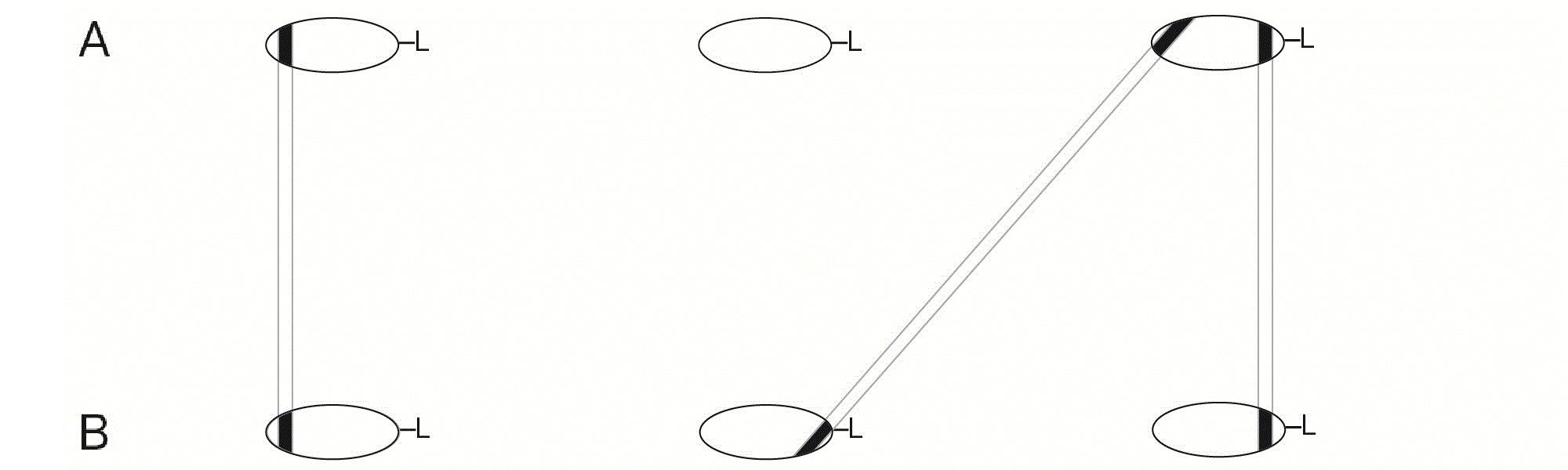}\end{overprint}
\begin{itemize}
\item <1->\textrm{Pick $\sai\cap\sbj$, }$\sai\cap\sbj\setminus p=\{\}$,
assign mass $d^{a_{i}},\, d_{b_{j}}$ from $ms^{a_{i}},\, ms_{b_{j}}$
\item <1->$d^{a_{i}}+d_{b_{j}}=c\lambda(\sai\cap\sbj)$
\item <1->If $(ms^{a_{i}}-c\lambda(\sai\cap\sbj))/(\lambda(\sai)+L^{a_{i}})\geq ms_{b_{j}}/(\lambda(\sbj)+L_{b_{j}})$
then $d_{b_{j}}=0$ 
\item <1->If $(ms_{b_{j}}-c\lambda(\sai\cap\sbj))/(\lambda(\sbj)+L_{b_{j}})\geq ms^{a_{i}}/(\lambda(\sai)+L^{a_{i}})$
then $d^{a_{i}}=0$ 
\item <1->otherwise \textrm{find $(ms^{a_{i}}-d^{a_{i}})/(\lambda(\sai)+L^{a_{i}})=(ms_{b_{j}}-d_{b_{j}})/(\lambda(\sbj)+L_{b_{j}})$}\end{itemize}
\begin{overprint}
\includegraphics<1->[scale=0.15]{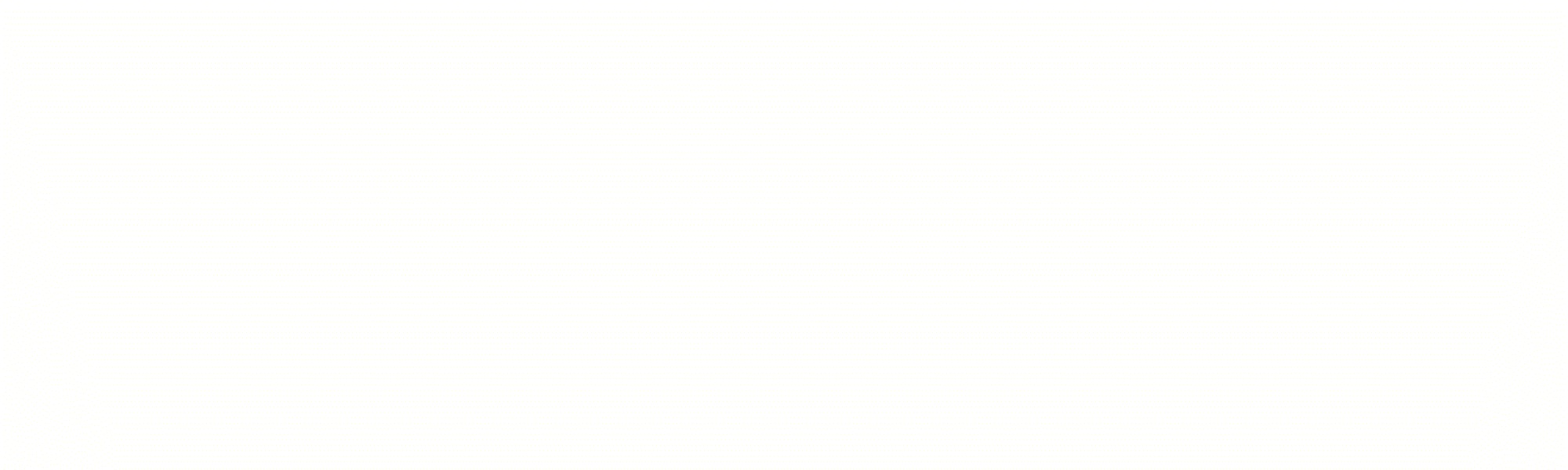}
\end{overprint}

\lyxframeend{}\lyxframe{Betting Decisions Step}

\begin{overprint}
\includegraphics<1->[scale=0.15]{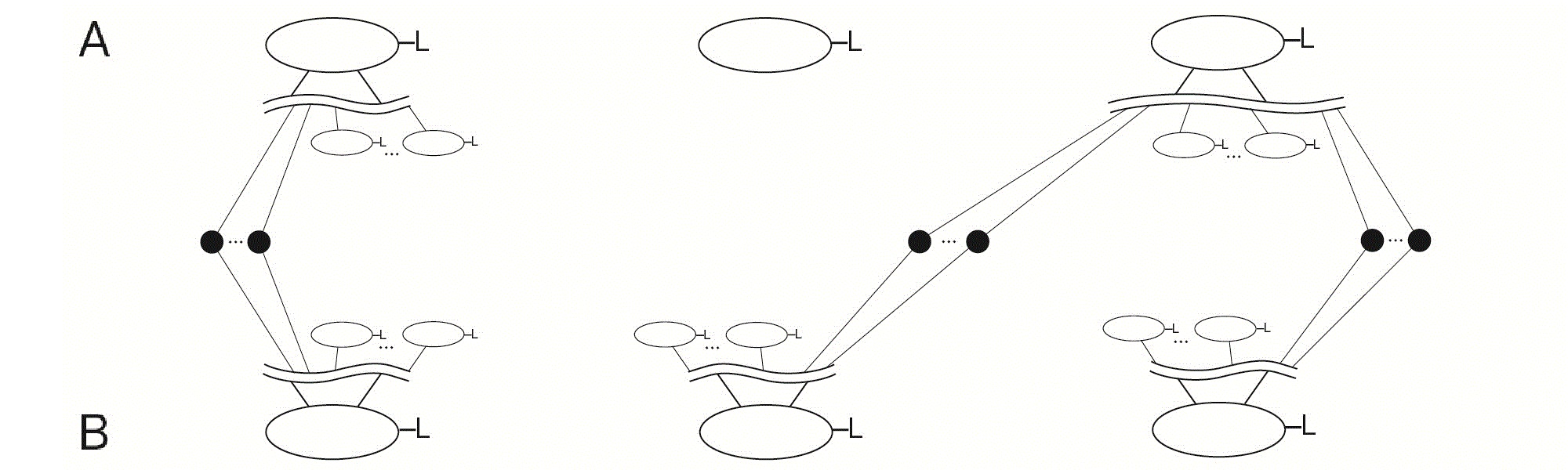}
\end{overprint}

\note[item]{TODO na slici napisi A,B i vidi zakaj jos uvijek pleshe}
\begin{itemize}
\item <1->Add the size of all words of a node that are extensions of $p$
to $L$
\item <1->For nodes that contain words $p^{\prime}$ s.t. $p\prec p^{\prime}$add
betting decisions:
\item <1->Leaf nodes with extensions of $p$ contain only one word, assign
mass from previous step, start another instance of the algorithm
\item <1->Distribute words unrelated to $p$ ammong remaining leaf nodes.\\Distribute
$L$ and the remainder of mass evenly ammong the remaining leaf nodes.
\item <1->\textrm{For the remaining nodes we have $\sai\cap\sbj=\{\}$
or\\$\lambda(s)\lambda(\sai\cap\sbj)$ between $(1\pm cs)(\lambda(\sai)+L^{a_{i}})(\lambda(\sbj)+L_{b_{j}})$}\end{itemize}
\begin{overprint}
\includegraphics<1->[scale=0.15]{prazna_vodoravna.jpg}
\end{overprint}

\lyxframeend{}\lyxframe{Proof Sketch}

\begin{columns}

\column{7cm}
\begin{itemize}
\item <1->$\Delta ms^{a_{i}}=ms^{a_{i}}-ms^{\prime a_{i}}$\\$f=\frac{\lambda(s)}{m}ms^{a_{i}}/(\lambda(\sai)+L^{a_{i}})$
\\$f^{\prime}=\frac{\lambda(s)}{m}ms^{\prime a_{i}}/(\lambda(\sai)+L^{a_{i}})$
\\$\Delta ms^{a_{i}}=\frac{m}{\lambda(s)}(f-f^{\prime})(\lambda(\sai)+L^{a_{i}})$
\item <1->$\Delta ms^{a_{i}}\leq c\lambda(\sai\cap\sbj)$\\$\lambda(\sai\cap\sbj)$
is close to $\frac{1}{\lambda(s)}(\lambda(\sai)+L^{a_{i}})(\lambda(\sbj)+L_{b_{j}})$
\item <1->$g=\frac{\lambda(s)}{m}ms_{b_{j}}/(\lambda(\sbj)+L_{b_{j}})$\\\textrm{$cL_{b_{j}}\geq(1-g)\frac{m}{\lambda(s)}(\lambda(\sbj)+L_{b_{j}})$\\Assume
that $g\leq f^{\prime}$}
\item <1->\textrm{$L_{b_{j}}\geq const.m^{2}(1-f^{\prime})(f-f^{\prime})$}
\item <1->$\underset{i=1}{\overset{z}{\sum}}2^{-l(p_{i})}\geq const.m^{2}/2\geq2^{-k}$
\end{itemize}

\column{3cm}
\begin{overprint}
\includegraphics<1->[scale=0.15]{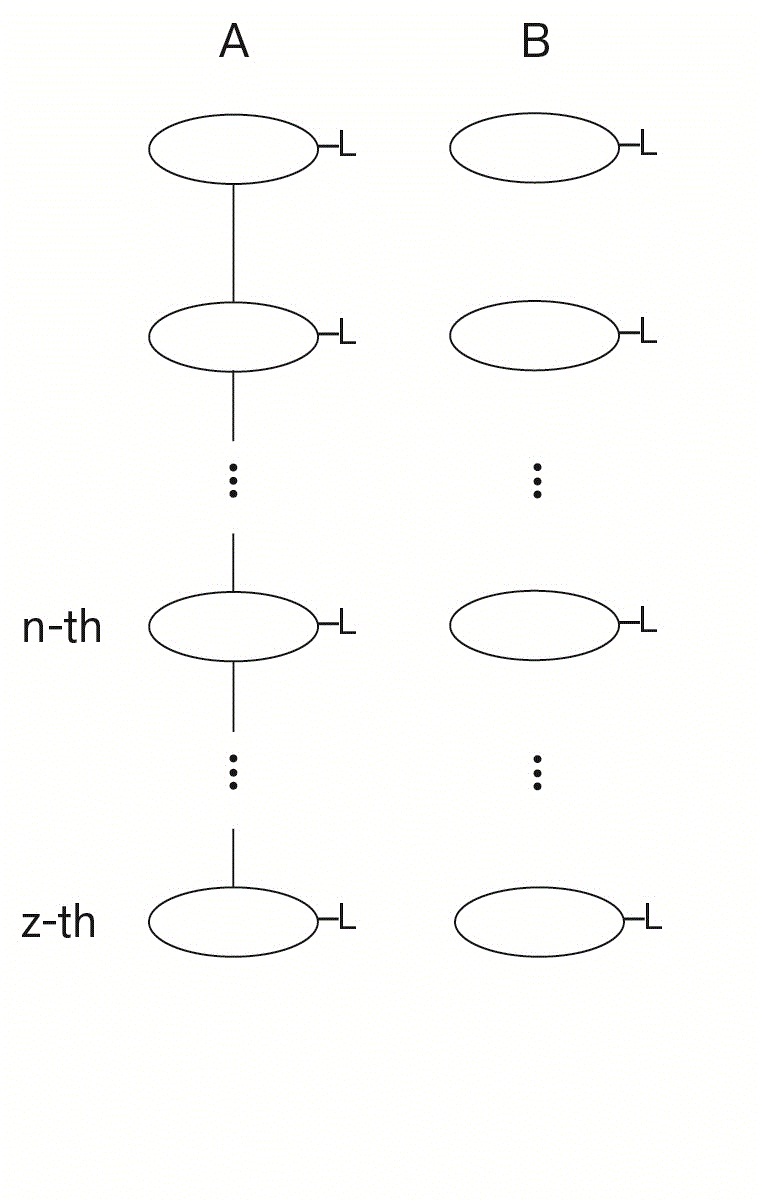}
\end{overprint}
\end{columns}

\lyxframeend{}

\appendix

\lyxframeend{}\section*{Appendix}

\lyxframeend{}\lyxframe{References}

\beamertemplatebookbibitems

\lyxframeend{}
\end{document}